# Depth evolution of YBa$_2$Cu$_3$O$_{7-\delta}$ ultrathin films probed by X-ray photoemission spectroscopy

*H. W. Seo[†,a], Q. Y. Chen [†, b] Paul. van der Heide[‡]* and Wei-Kan Chu

[†]*Department of Physics and Texas Center for Superconductivity, University of Houston, Houston, Texas, USA.*

[‡]*Department of Chemistry and Center for Materials Chemistry, University of Houston, Houston, Texas, USA.*

Abstract

X-ray photoemission spectroscopy has been used to investigate the depth dependent crystal structures and chemical compositions of sequentially chemical-etched YBa$_2$Cu$_3$O$_{7-\delta}$ (YBCO) ultrathin film superconductors. In the near-interface region the crystal structure is severely oxygen deficient and of tetragonal symmetry. We consider this a revelation of retarded oxygen diffusion into the O(1) sites during the post-deposition annealing in the presence of interface mismatch strain. Near the free surface, the oxygen-deficiency was much alleviated because of the partial strain relaxation and the crystal symmetry becomes orthorhombic. Compared with as-deposited films of equivalent thickness, which is less oxygen deficient and shows superconducting transition, the stripped-down near-interface layer exhibits no sign of superconductivity.

Oxygen diffusion, disordering and ordering of O(1) oxygen atoms on [-Cu(1)-O(1)-]$_n$ chains in YBa$_2$Cu$_3$O$_{7-\delta}$ (YBCO) lattice play a critical role in determining the physical properties of the superconductor.[1,2] Despite extensive studies, new anomalous phenomena continue to be discovered. Electron irradiation induced ordering of oxygen atoms by migrating from O(5) to O(1) sites is one case in point,[3] while revelation of superconductivity in tetragonal YBa$_2$Cu$_3$O$_{6.62}$ ($\delta=0.38$) is yet another.[4] These phenomena are considered anomalous as traditional wisdom believes, while nature favors randomness, for $\delta<0.65$ the compound is a superconductor of orthorhombic crystal structure. Though these new findings were made on bulk crystals, they have rekindled interests in the order-disorder issues pertinent to the general relationships among crystal symmetry, oxygen stoichiometry and superconductivity of the thin film materials. A relevant question is whether lattice-mismatch induced interface strain field influences the occupancy of the O(1) sites. This translates into whether or how the interface strain affects the diffusion of oxygen atoms into the YBCO thin films. As much of the crystal defects resides on the O(1) sites, in essence the curiosity is particularly about the equilibrated occupancy or vacancies of the O(1) sites after post-deposition oxygenation annealing. It is plausible that the gradient of chemical potential into the sample established by the oxygen partial pressure under the annealing conditions can be compensated by the present strain or stress field. This then would result in the demolition or enhancement of the driving force for chemical diffusion, depending on the stress-strain tensor field near the interface.

The near-interface chemistry and crystal structure are relevant not only to the understanding of superconductivity for ultrathin-film oxide superconductors, but also to practical applications. However, confounded effects of interface mismatch strain and variations in chemical compositions pose a great challenge to the proper characterizations of a coherent thin film. We believe the present XPS work is the first measurement that has effectively overcome some of the challenges for being able to pin-point the chemical effect separately. Granted, near-interface crystal structures have indeed been widely studied by use of X-ray diffraction (XRD).[5] XRD is, however, unable to distinguish whether the tetragonality is associated with oxygen deficiency or due to the forced lattice matching since both contribute independently to the ultimate crystal symmetry. Studies have also been carried out using Raman spectroscopy,[6] but the Raman approach is not reliable because of the superposed substrate signals. There is thus an incentive to seek new methods by which the chemical states and crystal symmetries over the entire film thickness can be probed directly.

Reported here is on using X-ray photoemission spectroscopy (XPS), assisted by sequential etching, to study the chemical stoichimetry and crystal symmetry evolutions of the YBCO ultrathin film. The experiment, albeit conducted in a retrospective manner, offers some glimpse of the interface mismatch effects on the oxygenation diffusion and the state of O(1)-defects. XPS provides unambiguous signature of chemical shifts of each atom at their distinctive valence states which are representative of the crystal symmetry as well as the oxygen stoichiometry. Its being a highly surface-specific technique due to the very small photoelectron escape depth (<5 nm) necessitates the layer by layer stripping. For thin film materials, though oxygen atoms on the $Cu(2)O(3)_2$ planes were already in place during the film deposition, most of the O(1) chain-oxygen atoms are yet to be introduced by post-deposition oxygen annealing. For properly prepared YBCO samples, the main variation of chemical composition is in the extent of oxygen deficiency $\delta$ contributed by the O(1) vacancies.

C-axis oriented YBCO thin films of 20 nm thick were used as the starting materials. At that thickness, the upper part of the film is already strain-relieved to some extent as proven by XRD. While why the strain-relaxation was not complete at 20 nm is still under investigation, this partially relaxed state allowed us to concurrently study the more strained bottom regions in comparison with the upper part where strain-relaxation had already commenced. Thin film deposition was conducted in ultra high vacuum by reactive thermal co-evaporation of the constituent atomic species on $CeO_2$ buffered YSZ substrates, typically in an oxygen ambient of $P_{O2} \approx 10^{-4}\text{-}10^{-2}$ mbar with substrate temperature $T_{Sub} \approx 680$ °C. The samples were then annealed in oxygen ambient at 500 °C under flowing oxygen with $P_{O2} \approx 100\text{-}300$ torr to maximize the rate of O(1)-site occupancy. The starting samples were then thinned down stepwise by wet-etching using a 0.1% Br-ethanol solution. We note that the more defective areas tended to etch faster. The root-mean-square (RMS) surface roughness of the as-grown 20 nm film is ~1 nm. After etching for 30, 60 and 90 seconds, the average film thicknesses become 15, 10 and 5 nm as their RMS roughness levels increase to 1.8, 2.8, and 3.6 nm, respectively; the depth resolutions are hence compromised accordingly.

The XPS spectra were collected on an XPS instrument from the Physical Electronics (Model 5700). Photoemission of electrons was produced using a monochromatic Al $K_\alpha$ x-ray source (1486.6 eV) operated at 350 W. The photoelectrons were allowed to pass through a hemispherical analyzer operated in fixed retardation ratio mode at 11.75 eV of pass energy. This results in an energy resolution of $\leq 0.51$ eV. All data were acquired with take-off angle $\alpha$ of the detected electrons ranging from $30^o$ to $70^o$. Note here $\alpha$ is defined with respect to intersection of the surface and the plane of incidence, viz. $\alpha = \pi/2 - \varphi$, where $\varphi$ is the polar angle defined with respect to the surface normal. The $30^o$-spectra would carry more of the signature of the chemical shifts of the surface reconstructed atoms or any other contaminants.

Shown in Fig. 1 are the representative XPS spectra collected for the O1s peaks over the three films of different thicknesses in which one sees distinct spectral features in the 527 to 530-eV range. These represent the emissions from the O1s states within the bulk oxide, while those in the 530 to 534-eV range arise from the surface bound atoms, primarily in the form of surface contaminants. These assignments are consistent with previous reports, which have attributed the lower binding energy shoulder at ~527.4 eV to the chain-oxygen atoms that are present in the orthorhombic phase, i.e. O(1) on the [-Cu(1)-O(1)-]$_n$ chains, as tabulated in table 1.[7,8] This is accompanied by a peak at ~528.2 eV, which represents the O(3)1s states for the Cu(2)-O(3)$_2$ planes of the orthorhombic phase. By comparison, this O(3)1s peak is present at ~528.9 eV for the tetragonal phase. The fact that the peaks at 527.4 eV and 528.2 eV diminish against that at 528.9-eV as the film thickness is decreased implies that the orthorhombic phase is not present near the interface where only the oxygen deficient tetragonal phase exists. This analysis of the O-1s states is mainly based on previously published experimental results[7,8] regarding YBCO crystals and films of different oxygen contents.

Comparison of spectra collected with α =30° to those with 70° also reveals the presence of the non-superconducting (tetragonal) phase of YBCO in the near-surface region (< 1 nm from the surface) in all films. This tetragonal phase is often found in XPS studies. It may well have arisen from the ultra high vacuum XPS measurements during which loosely bound chain oxygen atoms were lost by out-diffusion. Note the bottom 5-nm film largely has only one phase, viz. the tetragonal phase, since we see in the 5-nm sample no significant angular dependence for the bulk peak when comparing the dotted 30° and 70° data.

The tetragonal to orthorhombic evolution gains additional support from the Cu-2p spectra. This transition involves changes in Cu valence states as a result of oxygen content variations. When an oxygen vacancy comes to exist, two electrons are left behind which could then reduce specific metal ions in YBCO according to the following reactions in Kröger-Vink notation: [1,9]

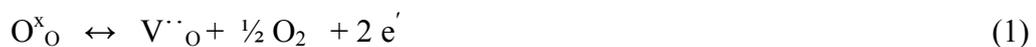

$$O^x_O \;\leftrightarrow\; V^{\cdot\cdot}_O + \tfrac{1}{2} O_2 \; + 2\, e^{'} \tag{1}$$

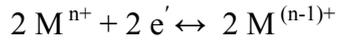

$$2\,M^{n+} + 2\,e^{'} \leftrightarrow 2\,M^{(n-1)+}$$

Copper is the most probable candidate for the valence change since it can exist in multiple oxidation states whereas Y and Ba do not. Since the main difference between orthorhombic and tetragonal YBCO is in the presence or absence of O(1) oxygen atoms on the [-Cu(1)-O(1)-]$_n$ chains, the neighboring Cu(1) sites would be the most directly affected. Indeed, when oxygen vacancies are formed, the $Cu^{2+}$ state is reduced to a $Cu^{1+}$ state.[10]

Divalent copper oxides exhibit a pair of spin-orbit split peaks of Cu-2p with binding energy $E_b$ in the range from 930 eV to 965 eV. The main core-level peaks are present at ~933.5 and 953.4eV. These are due to the well-screened $2p_{3/2}^{-1}d^{10}L^{-1}$ and $2p_{1/2}^{-1}d^{10}L^{-1}$ final states, respectively. Satellite peaks centered around 942 and 962 eV, due to unscreened $2p_{3/2}^{-1}d^9\,L$ and $2p_{1/2}^{-1}d^9\,L$ final states, where L refers to the ligand oxygen, are also present. The intensity ratios between the main and satellite lines for many other divalent copper oxide compounds, such as $La_2CuO_4$, $Nd_2CuO_4$, $Bi_2CuO_4$, $Ca_{0.85}Sr_{0.15}CuO_2$, and $SrCuO_2$, have been shown to be a good indicator for the degree of hybridization between the Cu $3d_{x2-y2}$ and O2p orbitals in the ground state,[12] as well as for the ionicity or the extent of hole-doping.[13,14] A configuration-interaction model [14] applied to interpret experimental results suggests that this ratio is indeed associated with the energy of charge transfer between Cu and O; the ligand becomes less electronegative when an oxygen is removed.

The above trends are also apparent on comparing the 5-nm film with their 10- and 15-nm counterparts, i.e. the 5-nm film essentially exhibits much weakened unscreened satellite lines at 942 and 962 eV, while the main peaks shift to lower binding energies, as shown in Fig.2-a. This again supports the near-interface oxygen deficiency scenario. As a cross examination, spectra were collected on a 20 nm sample subjected to vacuum annealing at an elevated temperature of 750 K to promote the production of chain-oxygen vacancies. For such sample, the suppression of the 942 and 962-eV satellite peaks and the 934.5 eV component of the main peak is indeed quite obvious as shown in Fig. 2-b. Here, in the $Cu^{+1}$ state, the unscreened peaks (satellites) are greatly reduced and the main core-level peaks should be at ~932.4 and 952.2 eV.[11]

To provide a further verification, analysis of the Ba-$3d_{5/2}$ spectral region, as shown in Fig. 3, reveals an increase in the binding energy of Ba relative to the bulk peak by 1.7 eV (Table 2) as the film is etched toward the interface, much the same as the Ba-$4d_{5/2}$ region (not shown here). The reasons for this increase are as yet unclear, but multiple-curve fittings of the Ba-$3d_{5/2}$ spectra (see insets in figs. 3) for the 5 nm film reveal a minimum of three different chemical bonding states. These peaks are assigned to the Ba bonding states for the tetragonal and orthorhombic phases and to those of the Ba atoms having been exposed to surface contamination. The reduction of the lower energy component closer to the interface for the 5 nm sample is consistent with the previous tetragonal-orthorhombic evolution picture in reference to bulk homogeneous samples of various of oxygen content. Although the precise interpretations for the Ba core-level spectra in the crystal symmetry evolution is still being debated, this result has been repeatedly confirmed under various experimental conditions.[15,16] Electrical transport measurements, which will be reported elsewhere,[17] also confirmed a consistent depression of superconducting transition temperature on each step of the etched film. The sample became semiconducting with no sign of superconductor-insulator transition down to 4 K when etched effectively to about 4-5 nm away from the interface. This is yet another support of oxygen deficiency and consequently the tetragonal symmetry since if the near-interface region were composed of more highly oxygenated orthorhombic phase compressed into a tetragonal lattice to fit the underlying square-lattice substrate, then the change in $T_c$ should in fact be quite small.[17,18] At the same time, the as-deposited utrathin film of comparable thickness demonstrate superconducting transition from 25-35 K. This led us to believe that interface strain had truly retarded the oxygen diffusion as strain field compromises the oxygen partial pressure gradient that drives the chemical diffusion.[1] Furthermore, for thicker films in which the strain is fully relaxed even in the near interface region, the thickness effect should be diminished and the in-diffusion of oxygen not retarded. This was exactly what was observed by Michaelis et al on samples which are thicker than 60 nm. [19]

In conclusion, we have shown the crystal-structural and stoichiometric evolutions for partially strain-relaxed YBCO films grown on $CeO_2$ buffered YSZ substrates using a retrospective wet chemical etching method. Measurement was done with X-ray photoelectron spectroscopy based on the consistent chemical shifts of O-1s, Cu-2p, and Ba-3d binding energies. We conclude that the 20-nm YBCO superconducting thin film exhibits orthorhombic crystal structure of higher chain oxygen content near the free surface region, but the orthorhombicity diminishes near the interface where the oxygen deficient tetragonal phase prevails. XPS is proven to be a useful technique to determine the crystal structure of YBCO thin film, much like what was established on the bulk crystals, based on the copper, oxygen and barium binding energies and charge states.

This work was supported in part by the State of Texas and the US Air Force Strategic Partnership for Research in Nanotechnology (SPRING) through the Texas Center for Superconductivity at the University of Houston, in part by National Science

Foundation under grant number DMR-0404542, and in part by the Department of Energy through grant number DE-FG02-05ER46208. Partial support by the Welch Foundation is also acknowledged.

a) e-mail: hseo@uh.edu.

b) Corresponding author, e-mail: Qchen@uh.edu.

References:

1. J.L. Routbort and S.J. Rothman, *J. Appl. Phys.* **1994**, 76, 5615.

2. Subhash L. Shinde, and David A. Rudman, *Interfaces in High Tc Superconducting Systems*: Springer-Verlag: New York, 1993.

3. H. W. Seo, Q. Y. Chen, M. N. Iliev, T. H. Johansen, N. Kolev, U. Welp, C. Wang, and W.-K. Chu, *Phys. Rev. B* **2005**, 72, 052501.

4. T. Frello et al, *Phys. Rev. B* **2003**, 67, 024509.

5. K. Kamamigaki et al, *J. Appl. Phys.* **1991**, 69, 3653; Tsai-Sheng Gau et al, *Appl. Phys. Lett.* **1994**, 65, 1720; W.J. Lin et al, *Appl. Phys. Lett.* **1998**, 72, 2966; W.J. Lin et al, *Appl. Phys. Lett.* **1998**, 72, 2995; A. Del Vecchio, M. F. De Riccardis, L. Tapfer, C. Camerlingo, and M. Russo, *J. Vac. Sci Technol.* A **2000**, 18, 802; A. Boffa and A. M. Cucolo, *Int. Jour. Mod. Phys. B* **2000**, 14, 2640-2645; B. Dam, J.M. Huijbregtse, and J.H. Rector, *Phys. Rev. B* **2002**, 65,064528.

6. P. Zhang, T. Haage, H. U. Habermeier, T. Ruf, and M. Cardona , *J. Appl. Phys.* **1996**, 80, 2935-2938.

7. R. P. Vasquez, B. D. Hunt, M. C. Foote and L J. Bajuk, and W. L. Olson, *Physica C* **1992**, 190, 249.

8. A. Hartmann, G.J. Russell, and K. N. R. Taylor, *Physica C* **1993**, 205, 78.

9. R. J. D. Tilley, *Defect Crystal Chemistry*; Blackie: New York, 1987.

10. J. M. Tranquada, S. M. Heald, A. R. Moodenbaugh, and Youwen Xu, *Phys. Rev. B* **1988**, 38, 8893.

11. S. Hufner, *Photoelectron Spectroscopy*; Springer-Verlag: New York, 2003. p. 111, 315, and 316.

12. J. M. Tranquads, and A.R. Moodenbaugh, *Phys. Rev. B* **1991**, 44, 5176.

13. T. Gourieux et al., *Phys. Rev. B* **1988**, 37, 7516.

14. J.-J. Yeh et al., *Phys. Rev. B* **1990,** 42, 8044.

15. R.P. Vasquez, *J. Electron Spectroscopy and Related Phenomena* **1993**, 66**,** 241.

16. M. Nagoshi, Y. Syono, M. Tachiki, and Y. Fukuda, *Phys. Rev. B* **1995**, 51, 9352.

17. U. Welp, et al., *Phys. Rev. Lev.* **1992**, 69, 2130.

18. H. W. Seo, Q. Y. Chen, Wolfgang Donner, Paul van der Heide, C. Wang and Wei-Kan Chu, (unpublished).

19. A. Michaelis, E.A. Irene, O. Auciello, and A.R. Krauss, *J. Appl. Phys.* **1998**, 83, 7736.

**Fig 1:**

**O1s XPS spectra of the 5, 10, and 15 nm films collected at a take-off angle of $30^o$ (upper panel, more surface sensitive data) and at $70^o$ (lower panel, less surface sensitive data).**

**Fig 2:**

**(a) Cu2p XPS spectra for 5, 10, and 15 nm films un-annealed with take-off angle at $30^o$ and $70^o$. (b) Those for a 20 nm film as deposited and after 750 K annealing; annealing results in loss of oxygen and reduced orthorhombicity.**

**Fig 3:**

**Ba3d XPS spectra of the 5, 10, and 15 nm films. The upper panel shows spectra collected at take-off angle at $30^o$ and $70^o$. Inset shows the multiple curve-fittings of the Ba-3d$_{5/2}$ peak (collected at $70^o$) for the 5 nm film  It reveals at least three different chemical bonding states representing the tetragonal and orthorhombic phases and surface contaminants (see Table 2).**

**Table 1: Reference peaks for the O1s states of YBCO.[5,6]**

**Table 2:  Reference peaks for the Ba3d and Ba4d states of YBCO.[13,14]**

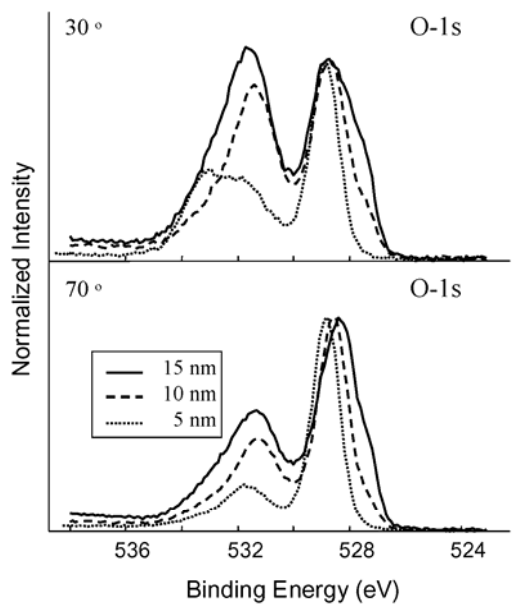

Fig 1

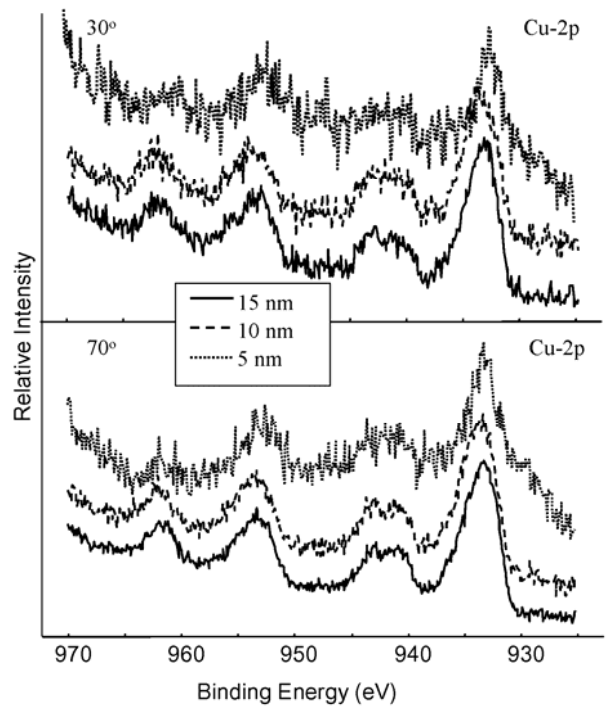

(a)

Fig 2

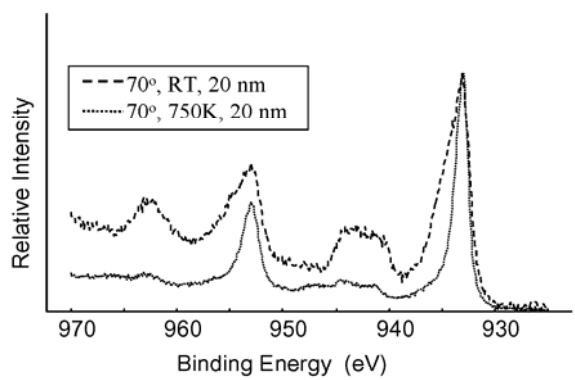

(b)

Fig 2

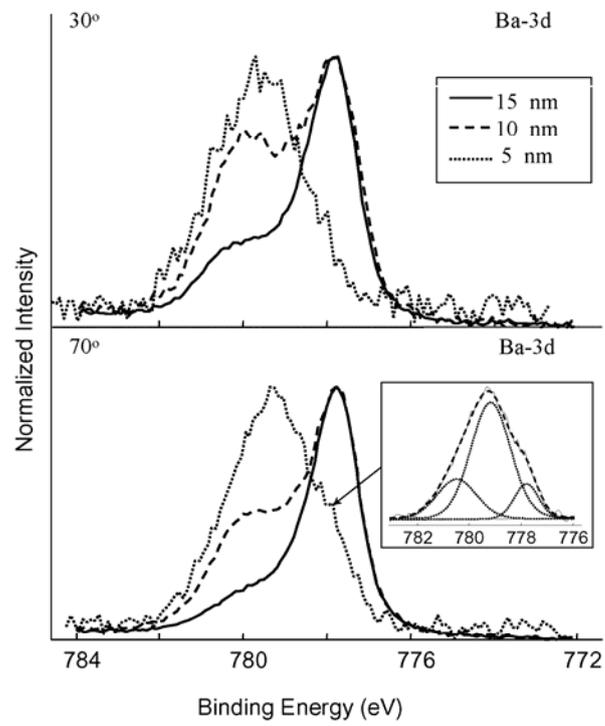

Fig 3

| | Binding Energy (Width) in eV | Description | Reference peak |
|---|---|---|---|
| O1s | 527.4 (0.8) | O ortho (chain) | 527.5 |
| | 528.2 (1.0) | O ortho (plane) | 528.2 |
| | 528.9 (1.0) | O tetra (plane) | 528.8, 528.7 |

Table 1

| | Binding Energy (Width) in eV | Description | Reference peak |
|---|---|---|---|
| $Ba-3d_{5/2}$ | 777.8 (1.2) | Ba ortho | 777.7 (film), 777.95 (poly) |
| | 779.5 (1.4) | Ba tetra | 778.6 (film), 779.28 (poly) |
| | 780.2 (1.7) | BaO Surface contaminants | - |

Table 2